# *LiveBo*: Empowering Non-Chinese Speaking Students through AI-Driven Real-Life Scenarios in Cantonese

Ka Yan FUNG
Institute of Special Needs and Inclusive Education
The Education University of Hong Kong
Hong Kong SAR, China
fkayan@eduhk.hk

Kwong Chiu FUNG
Hong Kong University of Science and Technology
Hong Kong SAR, China
kcfungag@connect.ust.hk

Yuxing TAO
Institute of Special Needs and Inclusive Education
The Education University of Hong Kong
Hong Kong SAR, China
ytao@eduhk.hK

Tze Leung Rick LUI
Institute of Special Needs and Inclusive Education
The Education University of Hong Kong
Hong Kong SAR, China
rtllui@eduhk.hk

Kuen Fung SIN
Institute of Special Needs and Inclusive Education
The Education University of Hong Kong
Hong Kong SAR, China
kfsin@eduhk.hk

## ABSTRACT

Language learning is a multifaceted process. Insufficient vocabulary can hinder communication and lead to demotivation. For non-Chinese speaking (NCS) students, learning Traditional Chinese (Cantonese) poses distinct challenges, particularly due to the complexity of converting spoken and written forms. To address this issue, this study examines the effectiveness of real-life scenario simulations integrated with interactive social robots in enhancing NCS student engagement and language acquisition. The research employs a quasi-experimental design involving NCS students who interact with an AI-driven, robot-assisted language learning system, *LiveBo*. The study aims to assess the impact of this innovative approach on active participation and motivation. Data are collected through proficiency tests, questionnaires and semi-structured interviews. Findings indicate that NCS students experience positive improvements in behavioural and emotional engagement, motivation and learning outcomes, highlighting the potential of integrating novel technologies in language education. We plan to compare with the control group in the future. This study highlights the significance of interactive and immersive learning experiences in promoting motivation and enhancing language acquisition among NCS students.

## 1 Introduction

Language learning is a complex process that involves the acquisition of vocabulary, grammar, sentence structure and also an understanding of the cultural context of the language [4]. For non-Chinese speaking (NCS) students, learning **Traditional Chinese (Cantonese)** presents unique challenges and opportunities, particularly given its spoken and written forms [7]. Insufficient and limited vocabulary can hinder NCS students' communication and writing skills, leading to monotonous dialogue. If this situation persists, students are likely to become demotivated. **Real-life scenario simulation** is an effective language teaching method [1]. By simulating authentic daily life situations, students repeatedly encounter vocabulary and sentence structures within meaningful contexts, thereby reinforcing their memory and enhancing their practical application skills [9]. This approach activates the brain's memory mechanisms, particularly episodic memory, enabling deeper retention compared to isolated vocabulary memorization. As students engage with language in context, they gain a sense of accomplishment [14], realizing that what they learn can be applied immediately in real-life situations. Additionally, the use of interactive **social robots** facilitates learning in a **low-pressure environment** [12]. Students can make mistakes during simulations without fear, while AI-driven robot-assisted systems provide immediate feedback and corrections [3]. This supportive environment encourages risk-taking with language, transforming the language acquisition process into an interactive experience [6]. Incorporating multisensory learning strategies, such as visual prompts and contextual sound effects, can further enrich the learning experience [8]. For example, robots can connect images and vocabulary. Although the possible benefits of real-life scenario simulations are recognized to enhance students' learning experiences, a significant gap remains in understanding how the integration of multisensory elements, interactive robots, and AI-driven real-life scenario simulations impact student engagement. While traditional paper-based learning is commonly used in schools, it often lacks an interactive and immersive environment that captivates students' curiosity. To this end, this work fills the gap by leveraging an AI-driven, robot-assisted interactive system with real-life scenario simulations. This study investigates the effectiveness of these elements in terms of enhancing engagement and performance. It provides additional supports to teachers to fine-tune their teaching approaches and motivate NCS students on their language learning journey.

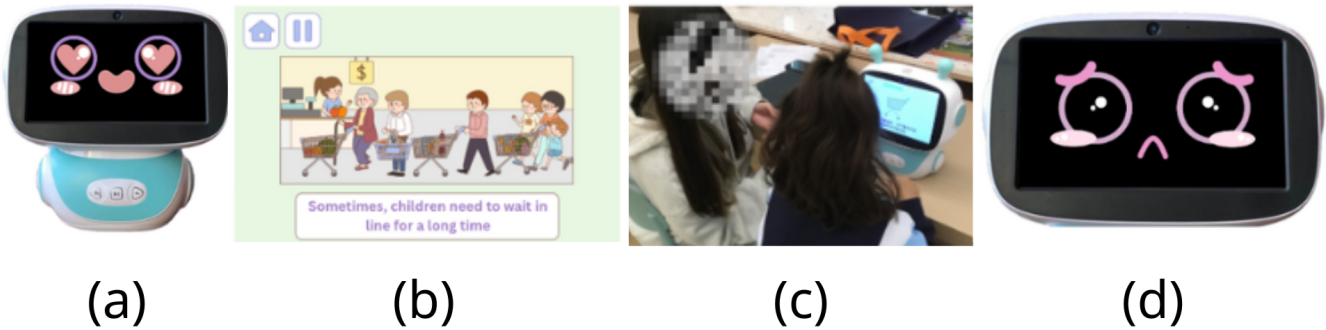

**Figure 1: The overview of the *LiveBo*: (a) is *LiveBo*, *Boon Boon*; (b) is a sample user-interface (UI); (c) was a NCS student interacting with *LiveBo*; (d) is a human-like animation.**

## 2 Study Design

The study follows a dual-phase framework involving (1) system development and (2) an initial exploratory study. We will discuss the flow and study in detail as follows.

### 2.1 Prototype Design

In this work, we developed an AI-driven real-life scenario simulation language learning system, *LiveBo*, (Figure 1) that features three common and local scenarios (i.e., supermarket, transportation and restaurant) presented on an interactive social robot, *Boon Boon* [5].

**Game Flow.** When students log into *LiveBo*, the learning process begins with a structured sequence of interactions. Students engage with the scenarios in a predetermined order, participating in real-life simulations and responding to Cantonese-related questions. Each question allows up to three attempts, with each attempt timed at 10 seconds. *LiveBo* employs AI technology, specifically Automatic Speech Recognition (ASR), to provide instant feedback. This feedback includes indicators of correctness, motivational quotes, and animations, all designed to enhance the interactive and personalized learning experience.

**Stories.** *LiveBo* features three common local scenarios: the supermarket, transportation, and restaurant. Each scenario is divided into two sections: local culture and vocabulary acquisition.

**Robot.** *Boon Boon* is equipped with a screen and four wheels. It can be programmed to mimic human-like reactions, such as dancing (i.e., rolling wheels) and singing (i.e., animations and sound effects).

**User Interface Design.** *LiveBo* incorporates multisensory elements and intuitive design to promote self-directed learning. It offers automatic responses to students and utilizes animations, sound effects, gestures, and movements to provide live interactions.

### 2.2 Exploratory Study

**Participant.** Ten NCS students participated in the study. Their average age was 8.20 years old (*SD* = 0.88 years).

**Procedure.** On Day 1, NCS students began with a pre-questionnaire prior to having the *LiveBo* learning. A total of three scenarios were presented, with one scenario each day, and each scenario took approximately 20 minutes to complete. An instructor provided technical support to two students. On Day 4, students completed a post-questionnaire and participated in an interview.

**Self-determination Theory and Interview.** Self-Determination Theory (SDT) is a psychological framework that fosters motivation and active participation in the learning process [11]. To assess these aspects, a set of 14 questions was set, focusing on behaviour, emotion, cognition, and intrinsic motivation, with three sub-questions under each category. For instance, one of the questions is, "When learning Cantonese through real-life scenarios using a robot, I feel interested." Furthermore, semi-structured interviews were conducted to collect student feedback, with a sample question being, "How do you feel about your experience learning Cantonese through real-life scenarios with a robot?"

**Inclusion Criteria.** To be eligible for participation, students were required to meet the following criteria: (1) Enrolling in Grades 1 to 3; (2) Not having any medical or physical disabilities that could impede their interaction with the robot or affect their communication abilities; (3) Being NCS students whose mother tongue was neither Traditional Chinese (Cantonese) nor Chinese (Mandarin); (4) Having at least one year of experience learning Cantonese. Furthermore, all participants were familiar with digital tools, including tablets. Parents provided informed consent for their children's participation before the study commenced. Participation was entirely voluntary, dependent on obtaining parental consent. The study was approved by the University's Institutional Review Board (IRB), and participants did not receive any financial compensation.

## 3 Data Analysis

The following section discusses data analysis on engagement, motivation, and learning performance.

NCS students demonstrated an improvement in **behavioural engagement** of 1.79%, $p=.78$ (Pre-test: $\bar{M} = 74.67$, $\sigma = 8.20$; post-test: $\bar{M} = 76.00$, $\sigma = 12.25$). In terms of **emotional engagement**, they showed an increase of 5.31%, $p=.57$ (Pre-test: $\bar{M} = 75.33$, $\sigma = 16.94$; post-test: $\bar{M} = 79.33$, $\sigma = 13.86$). Moreover, **intrinsic motivation** by 3.33%, $p=.68$ (Pre-test: $\bar{M} = 80.00$, $\sigma = 9.94$; post-test: $\bar{M} = 82.67$, $\sigma = 13.77$). However, there was a decrease in **cognitive engagement** by 4.27%, $p=.68$ (Pre-test: $\bar{M} = 78.00$, $\sigma = 16.04$; post-test: $\bar{M} = 74.67$, $\sigma = 19.06$). Regarding **learning proficiency**, NCS students improved by 20.59%, $p=.38$ (Pre-test: $\bar{M} = 56.67$, $\sigma = 29.61$; post-test: $\bar{M} = 68.33$, $\sigma = 27.72$).

## 4 Discussion

Implementing robot-assisted language training can enhance students' emotional engagement [10] and memory retention [13]. During scenario training, students showed a positive change in emotional involvement (+5.31%). The robot's immediate feedback provided a sense of satisfaction. For example, Student A was very enthusiastic and interacted with the robot when it expressed emotions, exclaiming, "I love this baby robot!"

Diep et al. [2] noted that learning a language also means embracing a new culture and lifestyle. Some NCS students, especially those from Urdu backgrounds, may find scenario training challenging due to their limited understanding of other languages. Cultural differences and language barriers can impede their participation. However, engaging with lively scenarios can help reduce their learning stress. Moreover, NCS students demonstrated a significant improvement in their overall scores (+20%) and retained vocabulary and phrases from the pre-test. When the same scenario was revisited during the post-test, many students excitedly responded with "Again!"

## 5 Conclusion and Future Work

NCS students demonstrate a substantial improvement with *LiveBo*. In the future, we plan to incorporate with a control group, increase sample size and develop a more comprehensive curriculum to ensure the reliability and generalizability of the study's findings.